# Topology Optimization through Differentiable Finite Element Solver

Liang Chen, Herman M.H. Shen
The Ohio State University

## Abstract

In this paper, a topology optimization framework utilizing automatic differentiation is presented as an efficient way for solving 2D density-based topology optimization problem by calculating gradients through the fully differentiable finite element solver. The optimization framework with the differentiable physics solver is proposed and tested on several classical topology optimization examples. The differentiable solver is implemented in Julia programming language and can be automatically differentiated in reverse mode to provide the pullback functions of every single operation. The entire end-to-end gradient information can be then backed up by utilizing chain rule. This framework incorporates a generator built from convolutional layers with a set of learnable parameters to propose new designs for every iteration. Since the whole process is differentiable, the parameters of the generator can be updated using any optimization algorithm given the gradient information from automatic differentiation. The proposed optimization framework is demonstrated on designing a half MBB beam and compared to the results with the ones from the efficient 88-line code. By only changing the objective function and the boundary conditions, it can run an optimization for designing a compliant mechanism, e.g. a force inverter where the output displacement is in the opposite direction of the input.

## 1  Introduction

Classical gradient-based optimization algorithms require at least first order gradient information to minimize/maximize the objective function. If the objective function can be written as a mathematical function, then the gradient can be derived analytically. For a complex physics simulator, such as finite element method (FEM), computational fluid dynamics (CFD), and multi-body dynamics (MBD), gradients are not readily available. To perform design optimization when the physics simulators are involved, a response surface [1] is constructed to approximate the behavior of the simulator. Response surface methodology and surrogate modeling use function fitting techniques to approximate the relationship between design variables (input) and the objective values (output). Once the response surface is constructed, gradient calculation and optimization algorithm (either gradient or non-gradient) are applied directly on the approximated function, which is much cheaper to evaluate. However, this approach may require a large number of precomputed data points to construct an accurate surface depending on the number of design variables. Therefore, generating large amount of data points arises as one of the bottlenecks of this method in practice due high computational cost of physics simulation. In addition, the surface may not generalize the behavior of the physics simulation accurately. This would cause the response surface to produce invalid solutions throughout the design space except at the precomputed data points. This result will be detrimental in terms of optimization, where the *best* solution must be found among all *true* solutions.

A possible reason response surface fails is its reliance on the function approximator to learn the underlying physical relationships using finite amount of data points. In theory, a function approximator is able to predict the relationship between any input and output according to universal approximation theorem [2, 3]. However, this usually requires enough capacity and complexity of the function, and sufficient number of data points to train the function. These data points are expensive to obtain and a design of experiments is performed at first to generate a set of combinations of design variables (data points) that can cover the whole design space uniformly. With the initial combination of design variables, a significant amount of effort must be placed towards computing the objective values of each data point through a physics simulator. At the end, only the design variables and corresponding objective values are paired for further analysis, i.e. function approximation and optimization. All the information during the physics simulation is discarded and wasted.

To make use of the computational effort during physical simulations, we turn to a newly emerged idea called differential programming [4, 5] which uses automatic differentiation to calculate the gradient of any mathematical/non-mathematical operations of a computer program. By using chain rule, one can back up the gradient information between any two variables during the computation process. As for the physics simulator, our goal is to use the automatic differentiation to compute the gradient between the objective value(output) and the design variable (input). In this paper, automatic differentiation is used to make a differentiable 2D structured finite element method. Then the differentiable physics engine is embedded inside an optimization loop to perform topology optimization. It is shown that this approach can achieve high quality results without providing analytical gradient formulas beforehand. Gradient values in the examples shown in Section 4 are backed up "automatically" by solving $KU = F$ during forward calculations. In addition, it is shown in Section 4.1 that the computation time is comparable to classical SIMP topology optimization method [6] such as the efficient 88-line code [7].

## 2  Automatic Differentiation

The idea of automatic differentiation uses derivative rules of known operations to calculate the derivatives of outputs with respect to inputs of a composite function $F: R^m \to R^n$, where $m$ and $n$ are the dimension of input and output, respectively. In general, the resulting derivatives form a $m \times n$ Jacobian matrix. When $m \gg n = 1$, the derivatives forms a gradient vector with length $m$. The automatic differentiation can be done in two ways: forward mode [8] or reverse mode differentiation [5]. In the following, ForwardDiff and ReverseDiff will be used for short. Usually the entire composite function $F$ has no simple derivative rule associated with. However, since it can be decomposed into other functions or mathematical operations where the derivatives are well defined, then the chain rule can be applied in order to propagate the derivative information either forward (ForwardDiff) or backward (ReverseDiff) through the computation graph of composite function $F$. The ForwardDiff has linear complexity with respect to the input dimension and therefore it is efficient for a function where the $m \ll n$. Further detail of the forward differentiation can be found in reference [8]. Conversely, the ReverseDiff is ideal for a function where $m \gg n$, and applied in this paper to calculate derivatives. ReverseDiff can be taken advantage of in this situation as the dimension of the input (design variables are densities of each element) is very large but the dimension of the output (objective is compliance or displacement criterion) is just one scalar value, such as overall compliance or displacement.

## 2.1 Reverse Mode Differentiation (ReverseDiff)

The name, ReverseDiff, comes from a registered Julia package called ReverseDiff.jl and additionally, has a library of well-defined rules for reverse mode automatic differentiation. The idea of ReverseDiff is related to adjoint method [5, 9, 10] and applied in many optimization problems where the information from the forward calculation of the objective function are reused to efficiently calculate the gradients. In structure optimization, the adjoint method corresponds to solve an adjoint equation $K\lambda = z$ [11, 12]. Since $K^{-1}$ is known when solving $KU = F$ during the forward pass, then the adjoint equation can be solved with a small computational cost without factorizing or calculating the inverse of matrix K. The ReverseDiff is also closely related to backpropagation [13] for training neural network in machine learning. The power of ReverseDiff is that it takes the automatic differentiation into a higher level, where any operation (mathematical or not mathematical) can be assigned with a derivate rule (called pullback function). Then with chain rule, we can combine the derivates of every single operation and compute the gradient from end to end of any black-box function, i.e. physics simulation engine or computer program.

To formulate the process for ReverseDiff, a composite function $F: y = f_n(f_{n-1}(\dots f_2(f_1(x_1))))$ is considered, where any intermediate step can be written as $x_{n+1} = f_n(x_n)$. A computation graph of the composite function is shown in Figure 1, for simplicity it is assumed that any intermediate function inside the composite function is single input and single output. The black arrows are the forward pass for function evaluation and the red arrows are reverse differentiation with chain rule been applied. The idea can be generalized to multiple inputs and outputs as well.

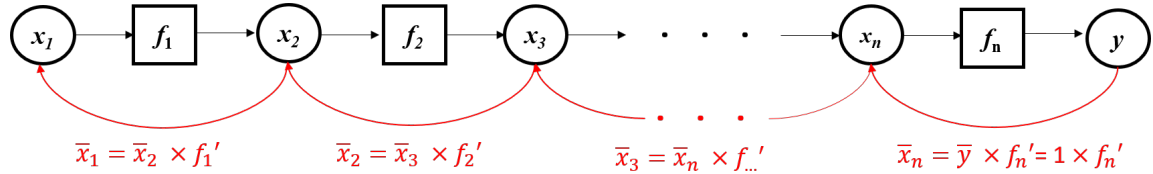

Figure 1 *A computation graph of a single input single output function*

To ReverseDiff of the function from end to end as shown in Figure 1, we will define and expand derivative using chain rule as:

$$\bar{x}_1 = \frac{dy}{dx_1} = \frac{dy}{dy} \times \frac{dy}{dx_n} \times \frac{dx_n}{dx_{n-1}} \times \dots \times \frac{dx_3}{dx_2} \times \frac{dx_2}{dx_1} = \bar{y} \times f'_n \times f'_{n-1} \times \dots \times f'_2 \times f'_1 \quad \textbf{Eq 2.1}$$

On the right side of the *Eq 2.1*, $\bar{y}$ is called seed and its value equals to 1. Starting from the last node, $y$, it is essentially calculating $\bar{x}_i = \frac{dy}{dx_i}$ as going backward along the computation graph. It can be shown that

$$\bar{x}_i = \frac{dy}{dx_{i+1}} \times \frac{dx_{i+1}}{dx_i} = \bar{x}_{i+1} \times f_i' \qquad \text{Eq 2.2}$$

In other words, for any function $x_{i+1} = f(x_i)$, the derivative of the input ($\bar{x}_i$) can be determined from the derivative of the output ($\bar{x}_{i+1}$) multiplied by $f_i'$. For ReverseDiff, *Eq 2.2* is known as the pullback function, as it pulls the derivate of the output backwards in order to calculate the derivate of the input. To evaluate the derivative using the pullback function, we need to know the output derivate as well as the value of $x_i$, thus a forward evaluation of all the intermediate values must be done and stored first before the reverse differentiation process. Theoretically, the computation time of the ReverseDiff is proportional to the number of outputs whereas for Forwarddiff, it scales linearly with the number of inputs. In practice, ReverseDiff also requires more overhead and memory to store the information during forward calculation. In the examples that are used in this paper, the input space dimension is on the order of $10^3 \sim 10^4$ but the output space dimension is only 1.

The pullback function is the rule that is needed to implement for every single operation during the forward calculation. For example, suppose the function we want to evaluate is $f: y = \sin(x^2)$. Then we need to write a generic pullback function for $b_1 = a_1^n$ as $\bar{a}_1 = \bar{b}_1 \times na_1^{n-1}$ and for $b_2 = \sin(a_2)$ as $\bar{a}_2 = \bar{b}_2 \times \cos(a_2)$. For the sake of demonstration, we can then combine these two pullback functions as one (will not do in practice as chain rule will take care of) it as:

$$B^f(\bar{y}) = \bar{y}\cos(x^2) \times 2x \qquad \text{Eq 2.3}$$

Notice that the pullback function of the exponent $x^2$ is known from calculus. However, we can also treat it as a multiplication of two numbers, the computation graph in this case is shown in Figure 2. The pullback function for multiplication of two real numbers, $y = x_1 \times x_2$, is simple:

$$B^f(\bar{y}) = \bar{y} \times x_2, \bar{y} \times x_1 \qquad \text{Eq 2.4}$$

Then the pullback function of the input $x$ in Figure 2 is simply the summation of the two terms in *Eq 2.3*, which is $\bar{y}(x_1 + x_2) = \bar{y} \times 2x$ when $x_1 = x_2 = x$.

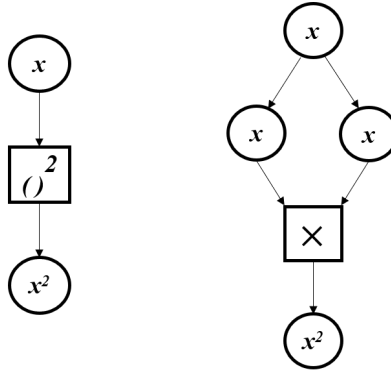
*Figure 2 Computation graphs of $x^2$ (left) and x*x (right)*

As can be seen from the example above, any high-level operations can be decomposed into a series of simple operations such as: addition, multiplication, sin()/cos(). Then the pullback function of the high-level operations can be always inferred from the known pullback functions with the chain rule. However, it will be a timesaver if the rules for some high-level operations can be defined directly. Just like the exponent function, it takes much longer computationally to convert it into a series of multiplications, while using a given rule from calculus like $d(x^n) = nx^{n-1}$ is much more efficient. In the structural topology examples that are demonstrated in Section 4.1, we will define a custom pullback rule for the backslash operator of the sparse matrix, which results in a much more efficient calculation of the gradient. The implementation details of this rule are discussed in Section 4.1.3.

**Programming Language**
Julia is used as the programming language. Julia has a great ecosystem for scientific computing and automatic differentiation. We use a Julia registered package called ChainRules.jl to define the custom pullback function of the finite element solver. Flux.jl was used to construct the neural network and for optimization.

## 3    Proposed Algorithm

### 3.1    Review of Solid Isotropic Material with Penalization (SIMP) Method

The SIMP method proposed by Bendsoe et.al. [6] established a procedure (Figure 3) for density-based topology optimization.

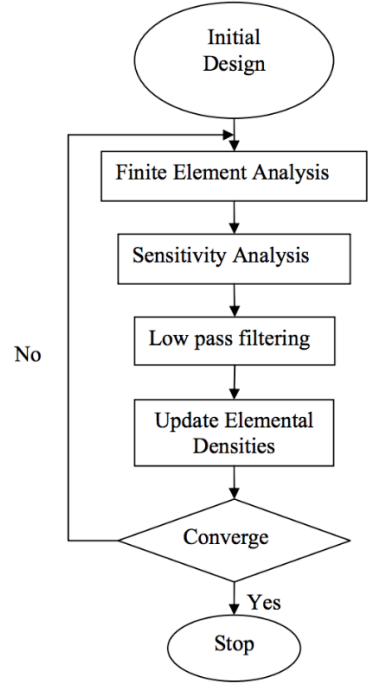

*Figure 3 SIMP method optimization process*

Given a predefined structured mesh in a fixed domain (Figure 5), the algorithm begins with an initial design by filling in a value ($x_e$) in each quadrilateral element. The value represents the material density of the element where 0 means a void and 1 means filled. A value between 0 and 1 is partially filled which does not really exist in real, it makes optimization easy but results in a blurry image of the design. Therefore, the author [6] proposed to add a penalization factor such that the stiffness is penalized if the density value is in between and push the element towards either void or completely filled.

The objective of the SIMP method is to minimize the compliance (*C*) of the design domain under fixed loadings and boundary conditions. The compliance defined in Eq 3.1 also described as total strain energy, is a measure of the overall displacement of a structure.

$$C(x) = U^T K U = \sum_{e=1}^{N} E_e(x_e) u_e^T K_0 u_e \qquad \text{Eq 3.1}$$

$$\frac{dC}{dx_e} = \frac{dE_e}{dx_e} u_e^T K_0 u_e \qquad \text{Eq 3.2}$$

$$x_e^{new} = \begin{cases} \max(0, x_e - m) & \text{if } x_e B_e^{\eta} \leq \max(0, x_e - m) \\ \min(1, x_e + m) & \text{if } x_e B_e^{\eta} \geq \min(1, x_e - m) ,\\ x_e B_e^{\eta} & \text{otherwise} \end{cases} \quad \text{where } B_e^{\eta} = \frac{-\frac{\partial C}{\partial x_e}}{\lambda \frac{\partial V}{\partial x_e}} \quad \text{Eq 3.3}$$

Inside the optimization loop, the density of each element ($x_e$) has to be updated in order to lower the compliance. For a gradient-based method, the gradient $dC/dx$ is calculated as *Eq 3.2* by taking derivative of *Eq 3.1*. Then a heuristic update rule (*Eq 3.3*) is crafted to ensure the optimality condition is satisfied for the design. In other popular density-based topology optimization techniques, such as level-set [14, 15] or bidirectional evolutionary optimization [16], an analytical formula of gradient, element density derivative or shape derivative, must be provided manually. In the proposed framework in Section 3.2, there is no need to provide such gradient information analytically as the gradients are calculated by reverse mode automatic differentiation.

## 3.2 Proposed Algorithm

In this work, we will perform the topology optimization using automatic differentiation for gradient calculation. The design variable update will be achieved by using a generator modeled as a convolutional neural network for 2D problem. A flow chart of this proposed design procedure is outlined in Figure 4.

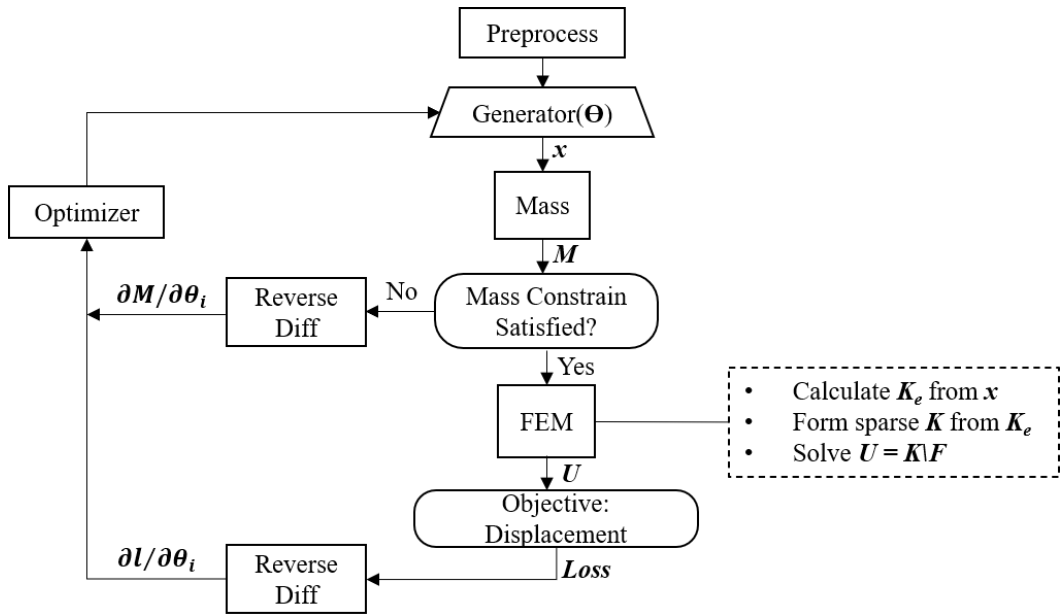

*Figure 4 Proposed optimization workflow with differentiable FEM solver*

Instead of providing a gradient equation (*Eq 3.2*) and update rule (*Eq 3.3*) manually, the gradient of every operation throughout the calculation is determined by using reverse differentiation (ReverseDiff). As for the objective function, the total strain energy, *Eq 3.1*, can be simply replaced by minimizing the displacement at the point of load. Therefore, the gradient is vector of $dL/dx_e$, where $L$ is a scalar value of the objective function. In the examples it is shown in Section 4 that the compliance/displacement is the objective function while a given total mass is an equality constrain that must be satisfied. Instead of providing an update rule (such as *Eq 3.3*), a component called a generator is added to propose new design (x) for every iteration. This generator (Figure 6) is essentially a neural network which is parameterized by a set of learnable weight coefficients (θ). The parameters of the network are adjusted simultaneously to generate better design while satisfying the mass constraint. Therefore, in the proposed algorithm, the gradients that guide the design approaching optimal are actually not $dl/dx_e$, but $dl/d\theta$ instead. For 1-D design case where the design variables have no spatial dependencies, the neural network is simply a multi-layer perceptron. While for 2D cases, the neural network architecture is composed of layers of convolutional operator which is good at learning patterns of 2-D images. Thus, the objective of this topology optimization becomes learning a set of parameters, θ, such that the generator is able to generate a 2D structure, *x*, such that it minimizes the displacement at the point of load while satisfy the mass constrain.

The optimizer component in Figure 4 is the algorithm to update the variables given the gradient information. In this work, ADAM (ADAptive Moments) [17] is used to handle the learning rate of and update the parameter θ of the generator in each iteration.

There is a preprocess component in Figure 4 used to set up the problem and initialize the variables that are non-differentiable. For example, the geometry (design domain), mesh (design domain discretization and element connectivity), boundary conditions, and loadings are fixed. Therefore, there is no need to make these parts differentiable. All other components, such as the generator, FEM solver, mass constraint, and the objective/loss functions have to be fully differentiable. Because each component is a composition of more specific functions (see the FEM component in Figure 4), that means the pullback rules have to be made for these functions explicitly or inferred from lower level functions.

## 4 Examples and Results

### 4.1 2D Density-based Topology Optimization

#### 4.1.1 Problem Statement

This is a well-known MBB beam (simply supported beam) for the benchmark test in topology optimization. The objective is to minimize the compliance of the beam subjected to a constant point load applied in the center. Due to symmetry about the vertical axis, the design domain (Figure 5) only include half of the original problem. As mentioned in Section 3.1, instead of minimizing the overall compliance, the objective function becomes to minimize the displacement at the point of the load. The equality constrain is such that the overall mass has to be kept as a constant fraction (0~1) of the maximum possible mass of the entire design domain.

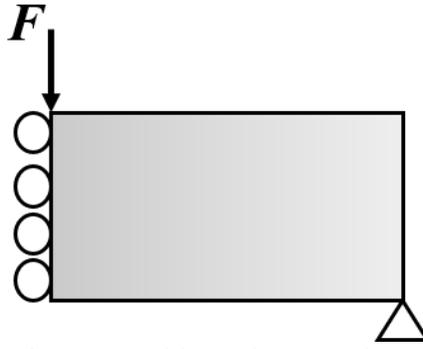

*Figure 5 Design domain and boundary condition of MBB beam*

### 4.1.2 Generator Architecture
The network architecture of the generator in Figure 4 is illustrated in Figure 6.

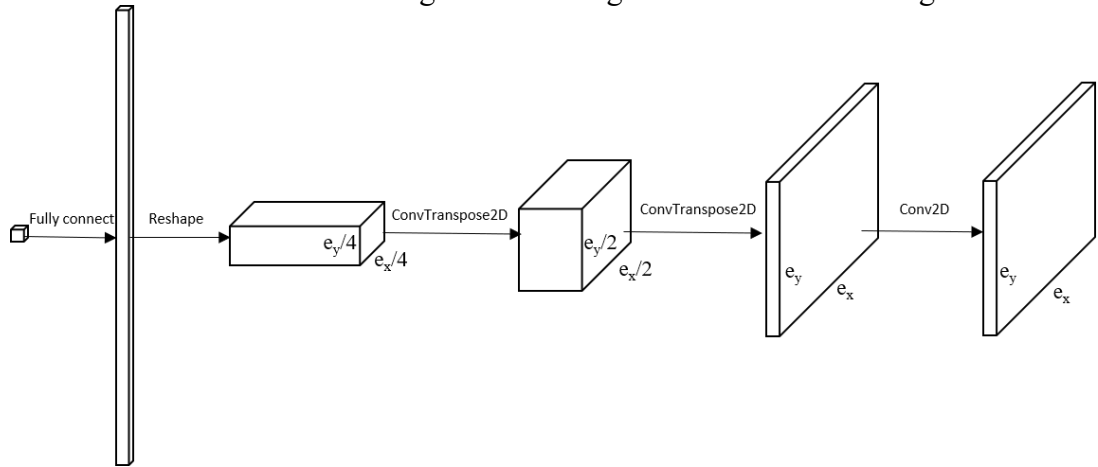

*Figure 6 Generator architecture*

Adapting the idea of generator from Generative Adversarial Network(GAN) [18], which is able to generate high quality images after training, the generator starts with a seed value (fixed or random) which followed by a fully connected layer. Then the fully connected layers are reshaped to a 3D array. The first two dimension relates to the width and height of the image and third dimension refers to channels. Then the subsequent components are to expand the width and height of feature layers but shrink the number of channels down to 1. At the end, the output will be a 2-D (3$^{rd}$ dimension is 1) image with correct size $e_x \times e_y$ based on the learnable parameter θ. The Reshape has no learnable parameters and the last Conv2D has fixed parameters. Only the parameters of the Full-connected and two ConvTranspose2D layers have learnable parameters θ. In Figure 6, the output size of ConvTranspose2D layer doubles every time. The second last layer has an activation function, *tanh*, which makes sure the output values are bounded. The last convolutional layer does not change the size of the input and its parameters are pre-determined and kept fixed. The purpose of the last layer is to average the density value around the neighborhood of each element to eliminate the checkerboard pattern. This is the filtering technique that was used in the SIMP method [7, 19].

The generator architecture can vary by adding more layers or new components. Instead of starting with a size of $e_x/4 \times e_y/4$, one can make this even smaller and add more channels at the beginning. As well, hyperparameters such as the kernel sizes, activation functions can be applied to any layers of the generator. However, in this paper, the architecture design is kept to be simple without experimenting too much with the hyperparameters of the architecture.

### 4.1.3 Differentiable FEM Solver

To construct a differentiable FEM solver, a standard forward calculation must be coded as illustrated in the dashed box in Figure 4. Then we need to make sure every operation in the forward calculation has a pullback function associated for ReverseDiff. Most elementary mathematical operations in linear algebra have pullback functions well-defined in Julia. However, operation for solving $KU = F$, where K is a sparse matrix, has not been defined. Therefore, it is important to write an efficient custom rule for the backslash operator $U = K\backslash F$. There are two parts associated with the backslash operation. The first part is to construct a sparse matrix $K = sparse(I,J,V)$, where $I$ and $J$ are the vectors of row and column indices for non-zero entries and $V$ is a vector of values associated with each entries. The pullback function of the operation is defined as $\bar{V}(\bar{K}) = NonZerosOf(\bar{K})$. The process is illustrated in Figure 7.

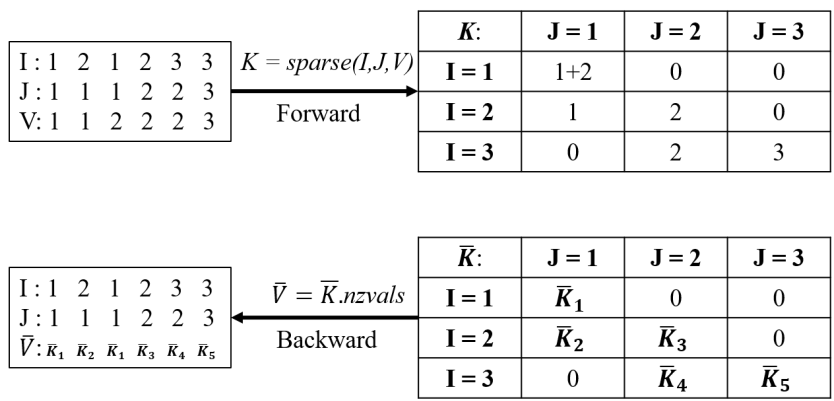

*Figure 7 Forward band Backward Calculation of sparse()*

The second part is for the backslash operation. For a symmetric dense matrix K, the pullback function of U = K\F can be written as:

$$\bar{K}(\bar{U}) = -\bar{F}U', where\ U = K\backslash F, \bar{F} = K\backslash \bar{U}$$

Eq 4.1

The $\bar{U}$ is the derivative of each element of U with respect to the downstream objective function. In *Eq 4.1*, it requires two backslash operations but in practice the factorization or inverse of K can be reused and only one expensive factorization is needed. This is the reason why the ReverseDiff can "automatically" calculate the gradient by only evaluating function in forward.

When K is sparse, *Eq 4.1* can be done efficiently by only using the terms in $\bar{F}$ and $U'$ that corresponds to the nonzero entries of the sparse matrix K. Otherwise, *Eq 4.1* will results in a dense K matrix which takes up significant memory.

### 4.1.4 MBB Beam Optimization Results.
Figure 8 shows a good convergence of MBB beam topology design within 100 iterations. The mass fraction is kept as 0.3. It is shown that the objective value in the vertical axis almost flats out at 40 iterations, where the design from the generator stays almost same afterwards.

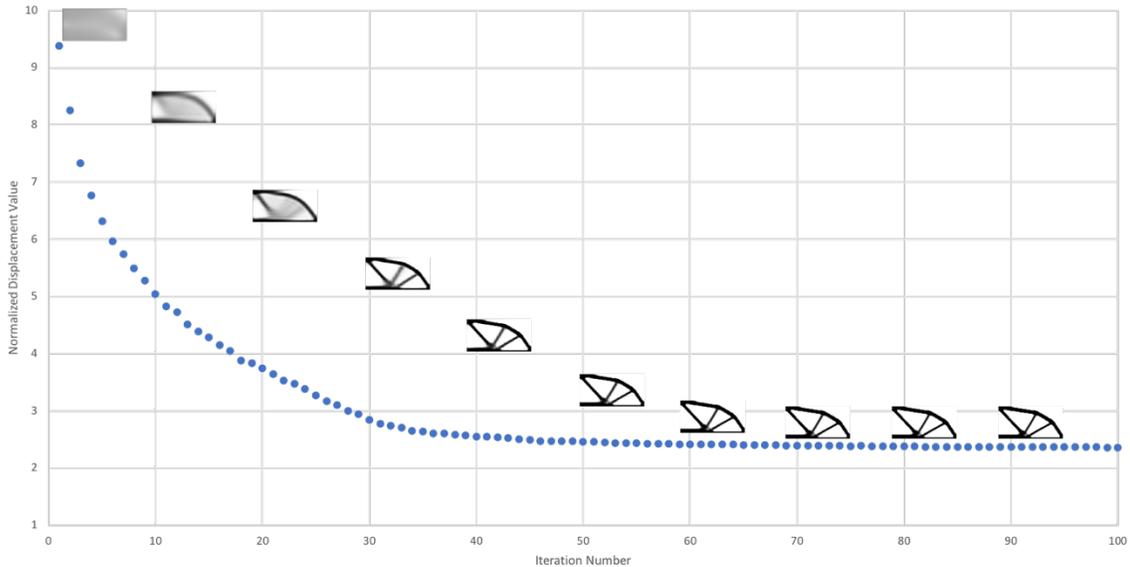

*Figure 8 Convergence of the objective function of proposed method*

When assembling the global stiffness matrix $K$ from the element density values $x$, the density value is penalized by using $x^p$, where $p \geq 1$. With a penalty factor, the checkerboard pattern is eliminated and a smooth boundary is created. Figure 9 shows side by side comparisons of the results from our proposed approach and SIMP 88-line code. All designs in Figure 9 are generated after 100 iterations for the MBB beam. The experiment ran combinations of three target mass fraction values and two penalty values for both methods. The number below each optimized design is the magnitude of the displacement at point of applied load. The proposed method works very well with low mass fraction design. When the mass fraction increases, however, the details of the design are hard to capture, and a higher penalty value is required to make the design clearer. Although the details of designs from two methods are different, both methods result in close displacement values. This means the optimized structures have equivalent overall stiffness given the mass fraction, loading and boundary conditions.

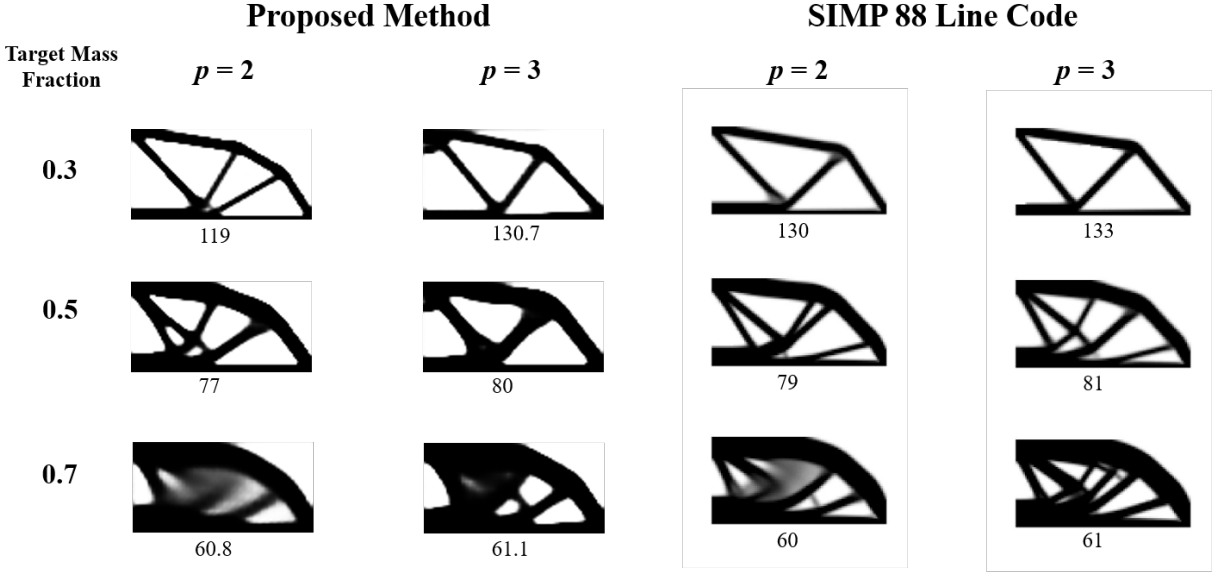
*Figure 9 Comparisons of results between proposed method and 88-line code*

Table 1 shows a computation time of proposed method on a 2015 Mac with 2.7GHz Intel i5 Dual Core and 8Gb memory. The time in the table is an average of 100 iterations. The actual time of each iteration varies due to the convergence of the mass constrain in the inner loop (Figure 4).

*Table 1 Computation time of proposed method for each iteration*

| Mesh Size | Time (seconds) per Iteration |
|---|---|
| 48*24 | 0.08 |
| 96*48 | 0.28 |
| 192*96 | 1.3 |

Figure 10 shows two more cases with different loading and boundary conditions (cantilever beam and bridge) on the density-based topology optimization using proposed design framework. The cantilever beam has fixed boundary conditions on left side and a tip load at midpoint of right side. The bridge design is simply supported at lower left and restricted motion in vertical direction at lower right. A point load is applied at midpoint at the bottom surface. The optimized structure using the proposed method does not look exactly like the one using SIMP 88-line code. However, they show similar trends, and for most cases in Figure 10 the displacement at point of applied load is less using the proposed method. But we cannot draw a conclusion that the proposed method in this paper is superior than 88-line code simply based on the slightly improvement on displacement values. This is because for density-based optimization, the stiffness is penalized for density between 0 and 1, therefore the overall stiffness values will vary even though the two designs look similar.

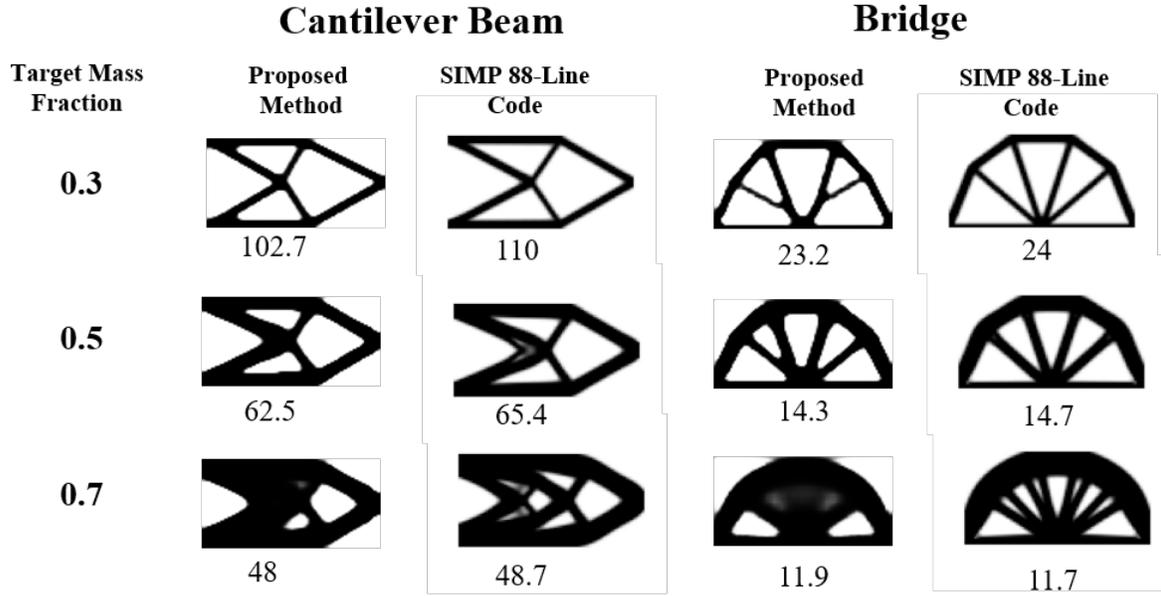

*Figure 10 Comparisons of Cantilever Beam (Left) and Bridge (Right) designs using proposed and SIMP 88-line method*

## 4.2 2D Compliant Mechanism Optimization

To demonstrate the flexibility of our proposed method, we use this approach to perform an optimization for compliant mechanism design. For a compliant mechanism, the structure is required to have low compliance at the point of applied load but high flexibility at some other part of the structure for desired motion or force output. A force inverter, for example, requires the displacement at point 2 to be in the opposite direction to the applied load at point 1.

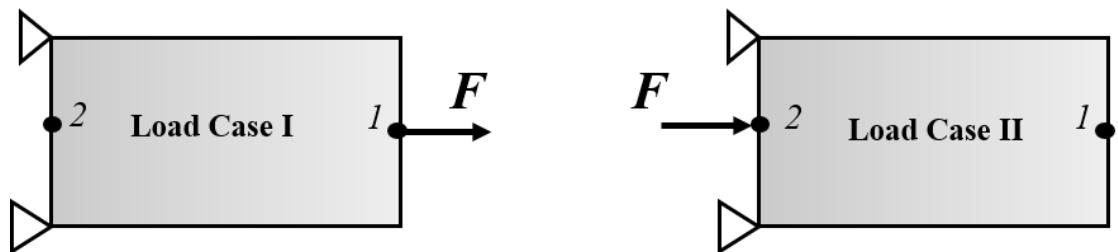

*Figure 11 Two load cases for optimizing a compliant mechanism*

Therefore, the objective function has to be formulated as two parts as shown in *Eq 4.2*.

$$L = \frac{U_2^I}{U_1^I} + w[U_2^{II} + U_1^I] \qquad \text{Eq 4.2}$$

The first term on the right-hand side is the geometry advantage, which is the ratio of the output displacement over input displacement for load case I. This term has to be minimized because the output displacement has to be negative which is opposite to the input force direction. Also, this term needs to be as negative as possible. However, with only this term in the objective function will result in intermediate density and very weak region around point 2. A second term is added to solve this problem, which measures the compliance of the overall structure, and we want to make sure the structure is stiff when load is applied at either point 1 or 2. The superscripts I and II denote two different loading cases (Figure 10). For case I, the force is applied at point 1 and displacements are recorded at both points 1 and 2. For case II, the force is applied at the point 2 and only the displacements at point 2 is recorded. Both displacements at the loading points of two cases have to be minimized in order to achieve low compliance. Therefore, for each iteration, two load cases, as opposed to one in the previous example, have to be run and the displacement value swill then be fed to the objective functions. The weight coefficient of the second term in *Eq 4.2*, *w*, has to be a small number (< 0.1), otherwise the design will be too stiff and result in very small geometry advantage.

The image on the left of Figure 11 is the optimized design of this force inverter where w = 0.01 and 0.3 volume ratio. It achieves a geometry advantage of 226/47

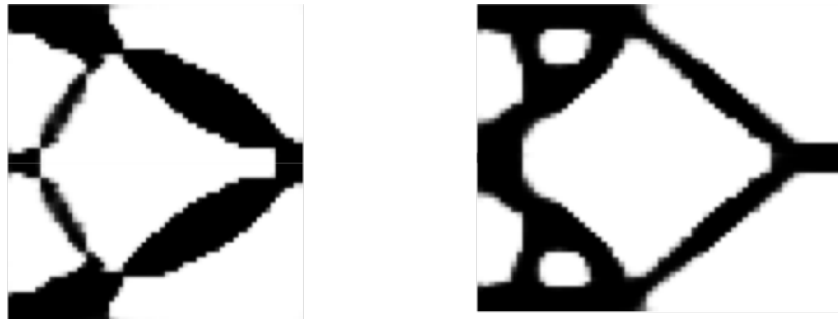

*Figure 12 Optimized force inverter with different objective functions. Left: use geometry advantage; Right: use target displacement*

The geometry advantage term in *Eq 4.2* can be replaced with $|U_2^I - U_{target}|$. This will make a design with the desired displacement at the output node. The image on the right of Figure 11 shows a design with desired target output displacement = -100. This design is much stiffer at point 2 due to the target displacement is smaller than the -226 of the design on the right of Figure 11

## 5   Conclusions

In this paper, a novel topology optimization framework with a differentiable 2D structured finite element solver is presented. Combining the differentiable solver with other differentiable layers, such as a convolutional generator, we can calculate the end-to-end gradient information of the whole computational graph, which can be used efficiently during optimization iterations. We demonstrated the proposed optimization framework on different test cases. Without explicitly specifying any analytical expression for gradient calculation and update rules, the generator after training is able to learn and produce promising results with appropriate objective functions. The generated optimized structure achieves the same level of overall stiffness as the well-known SIMP 88-line code. The proposed framework is much simpler to implement as only the forward calculations and basic derivative rules are required. Given the power of current computer technology and appropriate environment/ecosystem (e.g. Julia programming language) for implementation, the proposed framework can be applied to solve even more complex and larger engineering problems.

## Acknowledgement

I would like to thank Professor Simon Etter from National University of Singapore for his expert advice and help on Julia coding.